\documentclass[twocolumn,showpacs,amsmath,amssymb,prl,superscriptaddress]{revtex4-1}
\usepackage{graphicx}
\usepackage{bm}
\usepackage{dcolumn}

\begin{document}
\title{Electronic correlations stabilize the antiferromagnetic Mott state in Cs$_3$C$_{60}$}

\author{G. Giovannetti}
\affiliation{CNR-IOM-Democritos National Simulation Centre and International School
for Advanced Studies (SISSA), Via Bonomea 265, I-34136, Trieste, Italy}
\affiliation{Institute for Theoretical Solid State Physics, IFW-Dresden, PF 270116, 01171 Dresden, Germany}
\author{M. Capone}
\affiliation{CNR-IOM-Democritos National Simulation Centre and International School
for Advanced Studies (SISSA), Via Bonomea 265, I-34136, Trieste, Italy}

\date{\today}

\begin{abstract}
Cs$_3$C$_{60}$ in the A15 structure is an antiferromagnet at ambient pressure in contrast with other superconducting trivalent fullerides. Superconductivity is recovered under pressure and reaches the highest critical temperature of the family. Comparing density-functional calculations with generalized gradient approximation to the hybrid functional HSE, which includes a suitable component of exchange,  we establish that the antiferromagnetic state of Cs$_3$C$_{60}$ is not due to a Slater mechanism, and it is stabilized by electron correlation. HSE also reproduces the pressure-driven metalization. Our findings corroborate previous analyses suggesting that the properties of this compound can be understood as the result of the interplay between electron correlations and Jahn-Teller electron-phonon interaction.
\end{abstract}

\maketitle

Superconductivity in light-element materials entered in a new era after the recent observation of a critical temperature as high as 38K in Cs$_3$C$_{60}$\cite{Ganin} and the birth of a new family of organic superconductors based on aromatic molecules, as doped picene\cite{picene}, phenanthrene\cite{phenanthrene}, coronene\cite{coronene} and dibenzopentacene\cite{dibenz}. 
The unavoidable question about the ``pairing glue'' in the fullerides seems to have an answer at least for the standard members of the family, like K$_3$C$_{60}$ or Rb$_3$C$_{60}$. For these compounds there is convincing evidence of an electron-phonon driven superconductivity, in which Jahn-Teller-coupled intramolecular vibrations play the main role\cite{Gunnarsson_review}. Nonetheless, the anomalous properties of ``expanded'' fullerides with large intermolecular distance, like ammoniated compounds\cite{ammoniated} and most notably Cs$_3$C$_{60}$ suggest that labeling these materials as standard superconductors described by the Bardeen-Cooper-Schrieffer (BCS) theory is at least hazardous.

The anomalies are particularly clear in Cs$_3$C$_{60}$, in which the large Cs ions increase the lattice spacing in comparison with K- and Rb-doping. This compound is synthesized in two different crystal structure, the fcc which is common to most doped fullerides and a A15 structure, with a bipartite lattice\cite{Ganin2}. A15 Cs$3$C$_{60}$ at ambient pressure is an antiferromagnetic (AFM) insulator with $T_N$ = 46K and a spin-1/2 moment at each molecule\cite{Takabayashi}, as opposed to K$_3$C$_{60}$ and Rb$_3$C$_{60}$, which are superconductors in the same conditions. A superconducting state is recovered under external pressure of around 4Kbar. Even more interestingly, the critical temperature has a bell-shaped behavior as a function of pressure, with a maximum of $38K$ at P = 7KBar. 
The proximity between superconductivity and antiferromagnetism and the existence of a maximum remind of the phase diagram of several exotic superconductors, ranging from the high-temperature copper-oxides to heavy fermion materials, even if in the present case the bell-shaped curve is a function of lattice spacing instead of doping.
However, since the phase diagrams of these latter materials are believed to be dominated by strong electron-electron correlations and by the proximity to a Mott insulator, this similarity suggests that also the physics of Cs$_3$C$_{60}$ can be understood in the same framework.

The role of electron-electron correlations has been underlined well before Cs$_3$C$_{60}$ was known on the basis of estimates of the Coulomb interaction and the bandwidth\cite{gunnarsson_correlation,tosatti_correlation} In this light, the synthesis of Cs$_3$C$_{60}$ provides a system in which the correlation effects, already present in the standard fullerides, become predominant due to the reduced bandwidth associated to the larger distance between fullerene molecules. Once the role of correlations is realized, the AFM state of Cs$_3$C$_{60}$ can be interpreted as a Mott insulator with subsequent ordering of the Mott-localized electron spins. This picture has indeed been drawn using Dynamical Mean-Field Theory (DMFT)\cite{revdmft} for a three-band Hubbard model including the realistic electron-phonon interaction and the electron-electron correlation. These calculations have indeed predicted the phase diagram as a function of volume (or equivalently of the ratio $U/W$, where $U$ is the screened Coulomb repulsion and $W$ is the width of the conduction band), including the first-oder transition between a spin-1/2 AFM and the superconductor\cite{rmp}. Within the same scenario, the correlation effects also influence the superconducting phase enhancing the order parameter, despite its s-wave nature\cite{rmp,science}. The low-spin character of the insulator (the three electrons per molecule could also form a spin-3/2) is understood in terms of the Jahn-Teller interaction which favors the low-spin configuration\cite{low-spin,science}. 

However, given the bipartite character of the A15 lattice, we can not rule out a Slater character of the AFM state. In the Slater picture the AFM ordering does not involve localized magnetic moment, but it arises as a Fermi-surface instability of an underlying metallic state. A further confirmation of the correlated nature of the AFM state would also corroborate the idea that the same correlation effects are crucial to understand the superconducting state, while a Slater AFM would be more compatible with a standard BCS mechanism for the superconducting state. In order to further investigate the role of electron-electron interactions in the insulating state and to discriminate between a Mott and a Slater AFM state, in this work we tackle the physics of Cs$_3$C$_{60}$ in the A15 structure from a complementary perspective, using density-functional theory (DFT) to investigate the material specific properties and its bandstructure. 

We anticipate that our main conclusion will be that local and semilocal functionals (Local density approximation, LDA and generalized gradient approximation, GGA) are not able to properly describe the compound as an antiferromagnet, while using the hybrid functional introduced by Heyd, Scuseria and Ernzerhof (HSE)\cite{HSE} an AFM state is obtained, as well as the pressure-driven metalization. These findings confirm that a pure Slater picture does not hold for the AFM state, and that the inclusion of correlation effects is a necessary ingredient to properly describe the properties of expanded fullerides. 

Before entering the details of our calculations, we briefly recall the main features of the electronic structure. The bandstructure of doped fullerides reflects the molecular nature of the compound. The electronic bands correspond indeed  to the molecular orbitals of an isolated C$_{60}$ molecule. The most relevant bands for the physics of doped fullerides are those originating from the three $t_{1u}$ LUMOs of each $C_{60}$ molecule, which are expected to be populated by the three electrons donated by the alkali-metal atoms.  The molecular nature of the solid reflects into narrow bands, which lead to major influence of the sizeable  intramolecular Coulomb repulsion\cite{gunnarsson_correlation,Nomura} and also to a relevance of non-adiabatic effects\cite{pietronero} in the Jahn-Teller electron-phonon coupling which leads to the formation of Cooper pairs. 

The significant ratio $U/W \simeq 1.5$ is a serious threat for density-functional calculations based on LDA and GGA approximations for the exchange-correlation potential. Therefore most studies devoted to the role of correlations in fullerides have resorted to many-body approaches to solve model Hamiltonians, ranging from Quantum Monte Carlo\cite{gunnarsson_QMC} to Dynamical Mean-Field Theory\cite{gunnarsson_DMFT,science,rmp,pseudogap}. On the other hand, the development of hybrid functionals including correlation effects is extending the realm of materials which can be handled within DFT providing us with a powerful tool which can be used to assess the relevance of correlations in a material. In this work we follow this route and we compare the results of GGA with the Heyd-Scuseria-Ernzerhof functional (HSE)\cite{HSE} as implemented in VASP\cite{VASP} in order to establish the role of electronic correlations in A15 Cs$_3$C$_{60}$.

\begin{figure}
\includegraphics[width=1.0\columnwidth,angle=0]{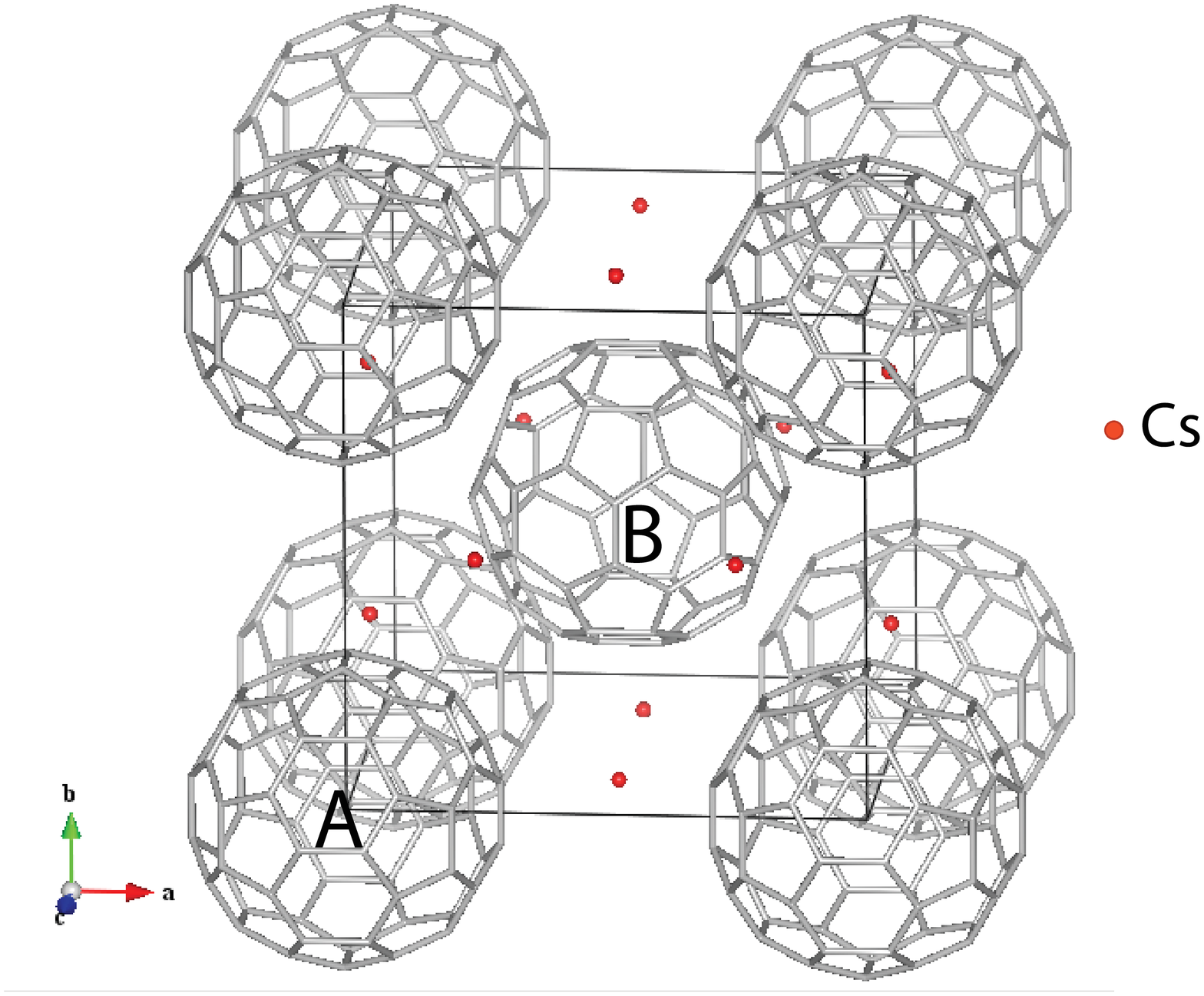}
\caption{(Color online) Schematic view of Cs$_3$C$_{60}$ unit cell with A15 structure showing the different orientation of the fulleride anions at the body center (B) and the origin of the coordinate system (A) and the position of the Cs cations}
\label{fig1}
\end{figure}

The electronic structure of Cs$_3$C$_{60}$ has been already investigated within LDA\cite{Darling,Nomura}. As a first step we complement the previous analyses using GGA in the Perdew-Burke-Ernzerhof (PBE)\cite{PBE} scheme. We use the projector augmented wave scheme\cite{PAW} and the cut-off for the plane-wave basis set was chosen as 400 eV and a 5$\times$5$\times$5 mesh was used for the Brillouin-zone sampling. Calculations are performed on the A15 structure of  Cs$_3$C$_{60}$ with Pm3n symmetry based on b.c.c. anion packing and refined at ambient pressure\cite{Ganin}. In this molecular packing the fulleride anion at the body center is rotated of 90 degrees around the [100] direction with respect to the anion at the origin as schematically shown in Fig. \ref{fig1}.

\begin{figure}
\includegraphics[width=.85\columnwidth,angle=0]{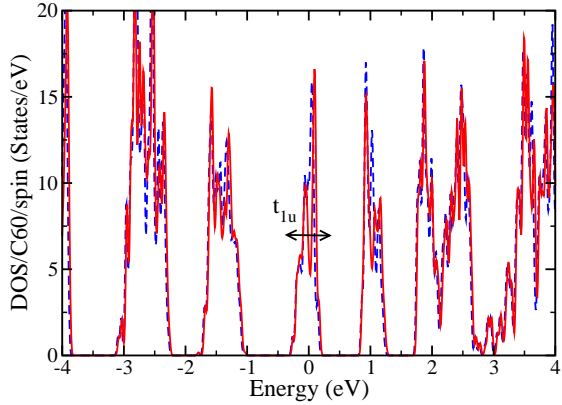}
\caption{(Color online) Density of states obtained in PBE for A15 Cs$_3$C$_{60}$ in a wide range of energies. The different bands arising from molecular orbitals are well visible and the $t_{1u}$ bands are emphasized. The blue dashed line is the nonmagnetic solution, while the red solid line is a metallic AFM solution}
\label{fig2}
\end{figure}

In Fig. \ref{fig2} we report the density of states for the t$_{1u}$ bands in the A15 structure obtained within PBE, which compare well with previous LDA calculations \cite{Darling,Nomura}. The three bands arising from the t$_{1u}$ LUMO are well separated from the rest of the bandstructure. The bandwidth is around 0.55 eV, confirming the expectation of a narrow-band solid. The crystal field preserves the degeneracy of the $t_{1u}$ orbitals and three electrons per fullerene molecule are donated by the Cs atoms leading to a half-filled manifold which can, at least in principle, be unstable toward antiferromagnetic (AFM) ordering with the two inequivalent sublattices associated to the two inequivalent fullerene molecule.

However, allowing for AFM ordering in the PBE scheme does not lead to the opening of an AFM gap. Imposing an AFM magnetic structure with opposite spins in the two sublattices A and B (shown in Fig.  \ref{fig1}), only a {\it metallic} AFM solution can be stabilized. Moreover this state does not lower the energy with respect to the nonmagnetic solution which remains the groundstate in this approximation. Therefore, despite the bipartite nature of the A15 structure, nesting is not sufficient to stabilize the AFM insulator which is experimentally observed and a pure Slater scenario can be ruled out.

\begin{figure}
\includegraphics[width=0.85\columnwidth,angle=0]{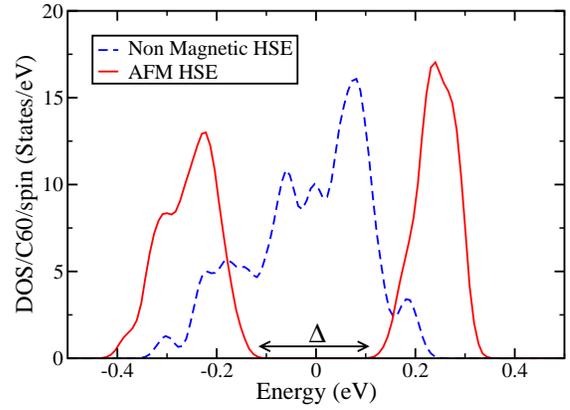}
\caption{(Color online) Density of states of the $t_{1u}$ valence bands obtained with HSE functional in the nonmagnetic (dashed blue line) and antiferromagnetic (solid red line). The arrow pinpoints the AFM gap $\Delta$}
\label{fig3}
\end{figure}

In light of the significant value of the intramolecular Coulomb interaction, electron correlations can be invoked to explain the discrepancy between PBE calculations and experiments. In order to overcome this limitation, we resort to the HSE functional which introduces interaction effects through a suitable fraction of the exact exchange term. The nonmagnetic solution does not introduce substantial differences with respect to PBE and only quantitative effects modify the bandstructure (see the blue dashed line in Fig. \ref{fig3}, where we focus only on the $t_{1u}$ bands). The situation changes dramatically for the AFM solution, whose density of states is plotted as a solid line in Fig. \ref{fig3}. As opposed to PBE, an antiferromagnetic state is instead stabilized within HSE. The AFM state is an insulator with a finite gap $\Delta$ of around 0.25 eV and an energy gain with respect to the metal of around 0.3 eV per unit cell. The ordered state has a magnetic moment of $\sim$ 1.4 $\mu_B$ per molecule, which is not far from the experimental value of $\sim$ 1 $\mu_B$ \cite{Takabayashi} which corresponds to a spin 1/2 state. 
The discrepancy can be understood within the scenario of Refs. \cite{low-spin,science,rmp}, where the low-spin state has been attributed to the Jahn-Teller interaction which favors low-spin configurations in the Mott state. In our HSE calculations we did not relax the ionic position because this is computationally too demanding, and we expect this would lead to a Jahn-Teller energy gain and consequently to a reduction of the molecular spin.

As the AFM state at ambient pressure is well reproduced we also performed a series of calculations aiming at describing the experimental pressure-driven metallization. Of course our DFT calculations do not allow for superconductivity and describe the experimental superconductor as a metal. The pressure-induced transition to bulk superconductivity takes place without change in crystal structure or symmetry \cite{Ganin}. The effect of pressure is therefore essentially a reduction of the lattice parameter, which can be easily taken into account within DFT. We performed calculations for the lattice parameters corresponding to 0, 5 and 10 kbar as determined in Ref. \cite{Ganin} without relaxing ionic positions. The reduced lattice spacing leads to a larger overlap between the t$_{1u}$ orbitals of different molecules and consequently to a larger bandwidth, which becomes 0.645 at 5 kbar and 0.685 at 10 kbar. The increased bandwidth leads to a reduced effect of correlations, which in turn modifies the energetic balance between the nonmagnetic and AFM states. At 5kbar the two solutions have the same energy within our accuracy, while at 10 kbar the metal is stable with an energetic gain of more than 0.3 eV per unit cell. This evolution remarkably mirrors the experimental findings and it strikingly confirms how a significant degree of  correlation is necessary to stabilize the AFM insulator. Interestingly, at 5 kbar the magnetic moment of the magnetic solution is still 1.34 $\mu_B$, only slightly below the ambient pressure value, despite the system is at the boundary between the two phases. This points towards a strongly first-order transition between an AFM with a finite magnetic moment and a metal which is expected to become superconducting. 

HSE is thus able to reproduce both the insulating behavior at ambient pressure and the metallic phase at finite pressure. Moreover, it provides an estimate of the transition which is close to the experimental result. Again, the discrepancy between our calculations and experiments can be attributed to the lack of ionic relaxation and the consequent underestimate of the Jahn-Teller energy gain. Including the Jahn-Teller interaction is indeed expected to reduce the magnetic moment and to slightly increases the energy of the AFM state. As a consequence, a calculation including the full Jahn-Teller distortion would provide a smaller critical pressure for the metalization. The ability of HSE to essentially reproduce the evolution of Cs$3$C$_{60}$ under pressure is a strong evidence of the accuracy of this approach to address strongly correlated molecular solids. This opens the path for a number of extensions which can bridge the gap between DFT calculations and, for example, DMFT solutions of a three-band Hubbard model.

The success of HSE for Cs$_3$C$_{60}$ can be used as an indirect indication of the accuracy of the same approach also for picene and for other organic superconductors, which all share similar values of $U/W$\cite{Nomura}. A HSE calculation for potassium-doped picene indeed confirms the relevance of correlations\cite{GGMCpicene}.
HSE has also been used to propose a positive role of electron correlations in determining the critical temperature of Ba$_{1−x}$ K$_x$ BiO$_3$ and β-HfNCl\cite{gabi_hse}.

In this paper we have demonstrated that HSE hybrid functional is able to reproduce the evolution of the properties of A15 Cs$_3$C$_{60}$ as a function of pressure, while PBE is not. In particular, HSE finds an AFM state at ambient pressure and an insulator-to-metal transition under external pressure, while PBE always gives a metallic solution. The calculations give a magnetic moment which is slightly larger than the experimental value, a difference which can be ascribed to the lack of ionic relaxation in our calculations. Only the full distortion of the molecule would allow for a full Jahn-Teller energy gain and a spin-1/2 local moment. 
Our results strengthen the idea that electronic correlations are at the basis of the properties of Cs$_3$C$_{60}$. This provides an indirect confirmation of  the strongly correlated superconductivy scenario\cite{science,rmp}. The similarity between the present results and previous calculations for potassium-doped picene suggest that the aromatic superconductors may share many similarities with doped fullerides. A further experimental test of the Mott nature of the AFM state can be provided by spectroscopic methods. It has indeed been shown theoretically that a three-dimensional Mott antiferromagnet should display a peculiar series of spin-polaron satellites both in the photoemission spectra\cite{sangiovanni} and in the optical conductivity, in which equally spaced spin-polaron peaks are  predicted to appear above the optical gap\cite{Taranto}.

We acknowledge useful discussions with M. Fabrizio and E. Tosatti and financial support by European Research Council under FP7/ERC Starting Independent Research Grant ``SUPERBAD" (Grant Agreement n. 240524). Computational support from CINECA and CASPUR is gratefully acknowledged.



\begin{thebibliography}{99}
\bibitem{Ganin} A.Y. Ganin {\it et al.}, Nature Mater. {\bf 7}, 367 (2008)

\bibitem{picene} R. Mitsuhashi {\it et al.}, Nature {\bf 464}, 76 (2010)

\bibitem{phenanthrene} X.F. Wang,	 R.H. Liu,	 Z. Gui,	 Y.L. Xie,	 Y.J. Yan,	 J.J. Ying,	 X.G. Luo, and X.H. Chen, Nat. Comm. {\bf 2}, 507 (2011)

\bibitem{coronene} Y. Kubozono {\it et al.}, Phys. Chem. Chem. Phys. {\bf 13}, 16476 (2011)

\bibitem{dibenz} M. Xu, T. Cao, D, Wang, Y. Wu, H. Yang, X. Dong, J. He, F. Li, and G. F. Chen, arXiv:1111.0820

\bibitem{Gunnarsson_review} O. Gunnarsson, Rev. Mod. Phys. {\bf 69}, 575 (1997)

\bibitem{ammoniated} M. Ricc\`o, G. Fumera, T. Shiroka, O. Ligabue, C. Bucci, and F. Bolzoni, Phys. Rev. B {\bf 68}, 035102 (2003)

\bibitem{Ganin2} A.Y. Ganin {\it et al.}, Nature {\bf 466}, 221 (2010); , Y. Ihara, H. Alloul, P. Wzietek, D. Pontiroli, M. Mazzani, and M. Ricc\`o,  Phys. Rev. Lett. {\bf 104}, 256402 (2010)


\bibitem{Takabayashi} Y.~Takabayashi {\it et al.}, Science {\bf 323}, 1585 (2009)

\bibitem{gunnarsson_correlation} V.P. Antropov, O. Gunnarsson and O. Jepsen, Phys. Rev. B {\bf 46}, 13647 (1992)

\bibitem{tosatti_correlation} M. Fabrizio and E. Tosatti, Phys. Rev. B {\bf 55}, 13465 (1997)

\bibitem{revdmft} A. Georges, G. Kotliar, W. Krauth, and M.J. Rozenberg, Rev. Mod. Phys. {\bf 68}, 13 (1996)

\bibitem{rmp} M. Capone, M. Fabrizio, C. Castellani, and E. Tosatti, Rev. Mod. Phys. {\bf 81}, 943 (2009)

\bibitem{science} M. Capone, M. Fabrizio, C. Castellani, and E. Tosatti, Science {\bf 296}, 2364 (2002).

\bibitem{low-spin} M.~Capone, M.~Fabrizio, P.~Giannozzi, and E.~Tosatti, Phys. Rev. B {\bf 62}, 7619 (2000)

\bibitem{HSE}  J.Heyd, G. E. Scuseria, and M. Ernzerhof, J. Chem. Phys. {\bf 118}, 8207 (2003); J. Heyd and G. E. Scuseria, J. Chem. Phys. 124, 219906 (2006)

\bibitem{Nomura}  Y. Nomura, K. Nakamura, and R. Arita, arXiv:1112.3483

\bibitem{pietronero} L. Pietronero and S. Str\"assler, Europhys. Lett. {\bf 18}, 627 (1992); L. Pietronero and E. Cappelluti, Low Temperature Physics {\bf 32}, 340 (2006)

\bibitem{gunnarsson_QMC} O. Gunnarsson, E. Koch, and R. M. Martin, Phys. Rev. B {\bf 54}, R11026 (1996); E. Koch, O. Gunnarsson, and R. M. Martin, Phys. Rev. Lett. {\bf 83}, 620 (1999)

\bibitem{gunnarsson_DMFT} J.~E.~Han, E. Koch, and O. Gunnarsson, Phys. Rev. Lett. {\bf 84}, 1276 (2000)

\bibitem{pseudogap} M. Capone, M. Fabrizio, C. Castellani, and E. Tosatti, Phys. Rev. Lett.  {\bf 93}, 047001 (2004);  M. Schir\`o, M. Capone, M. Fabrizio, and C. Castellani, Phys. Rev. B {\bf 77}, 104522 (2008)

\bibitem{VASP} G. Kresse and J. Furthmuller, Phys. Rev. B {\bf 54}, 11 169 (1996); G. Kresse and J. Furthmuller, Comput. Mater. Sci. {\bf 6}, 15 (1996)

\bibitem{Darling} G.R. Darling, A.Y. Ganin, M.J. Rosseinsky, Y. Takabayashi, and K. Prassides, Phys. Rev. Lett. {\bf 101}, 136404 (2008)

\bibitem{PBE} J.P. Perdew, K. Burke, and M. Ernzerhof, Phys. Rev. Lett. {\bf 77}, 3865 (1996)

\bibitem{PAW} G. Kresse and D. Joubert, Phys. Rev. B {\bf 59}, 1758 (1999).

\bibitem{GGMCpicene} G. Giovannetti and M. Capone, Phys. Rev. B {\bf 83}, 134508 (2011)

\bibitem{gabi_hse} Z. P. Yin, A. Kutepov, G. Kotliar, arXiv:1110.5751

\bibitem{sangiovanni} G.~Sangiovanni {\it et al.}, Phys. Rev. B {\bf 73}, 205121 (2006)

\bibitem{Taranto} C. Taranto, G. Sangiovanni, K. Held, M. Capone, A. Georges, and A. Toschi, Phys. Rev. B {\bf 85}, 085124 (2012)


\end{thebibliography}
\end{document}